\newcounter{myctr}
\def\myitem{\refstepcounter{myctr}\bibfont\noindent\ifnum\themyctr>9\else\phantom{0}\fi\hangindent17pt\themyctr.\enskip}

\documentclass{ws-ijqi}
\usepackage{hyperref}
\usepackage{graphicx}  % needed for figures
\usepackage{dcolumn}   % needed for some tables
\usepackage{bm}        % for math
\usepackage{amssymb}   % for math
\usepackage{amsmath}
\usepackage{dsfont}
\usepackage{cancel}
\usepackage{empheq}
\usepackage{graphicx,color}
\usepackage{cleveref}
\usepackage[super,sort,compress]{cite}
\newcommand{\unit}{1\!\!1}
\def\ket#1{| \,#1\, \rangle}
\def\bra#1{\langle \,#1\, |}
\begin{document}

%%%%%%%%%%%%%%%%%%%%% Publisher's Area please ignore %%%%%%%%%%%%%%
\catchline{}{}{}{}{}
%%%%%%%%%%%%%%%%%%%%%%%%%%%%%%%%%%%%%%%%%%%%%%%%%%%%%%%%%%%%%%%%%%%

\title{ENTROPY AND GEOMETRY OF QUANTUM STATES}

\author{KUMAR SHIVAM}

\address{Theoretical Physics, Raman Research Institute, Address\\
C.V.Raman Avenue, Sadashivnagar, Bangalore-560080
India\\
kshivam@rri.res.in}

\author{ANIRUDH REDDY}

\address{Theoretical Physics, Raman Research Institute, Address\\
	C.V.Raman Avenue, Sadashivnagar, Bangalore-560080
	India\\
	anirudhr@rri.res.in}

\author{JOSEPH SAMUEL}

\address{Theoretical Physics, Raman Research Institute, Address\\
	C.V.Raman Avenue, Sadashivnagar, Bangalore-560080
	India\\
	sam@rri.res.in}

\author{SUPURNA SINHA}

\address{Theoretical Physics, Raman Research Institute, Address\\
	C.V.Raman Avenue, Sadashivnagar, Bangalore-560080
	India\\
	supurna@rri.res.in}
\maketitle

%\begin{history}
%\received{11 September 2017}
%\revised{Day Month Year}
%\accepted{Day Month Year}
%\comby{(xxxxxxxxxx)}
%\end{history}

\begin{abstract}
We compare the roles of the Bures-Helstrom (BH)
and Bogoliubov-Kubo-Mori (BKM)
metrics in the subject
of quantum information geometry. We note that there are two
limits involved in state discrimination, which we call the
``thermodynamic'' limit (of $N$, the number of realizations going to infinity)
and the infinitesimal limit (of the separation of states tending to zero).
We show that these two limits do not commute in the quantum case. Taking
the infinitesimal limit first leads to the BH metric and the corresponding
Cram\'{e}r-Rao bound, which is widely accepted in this subject. Taking limits
in the opposite order leads to the BKM metric, which results in a weaker
Cram\'{e}r-Rao bound. This lack of commutation of limits is a purely quantum
phenomenon arising from quantum entanglement.  We can exploit this phenomenon
to gain a quantum advantage in state discrimination and get around the limitation imposed by the Bures-Helstrom Cram\'{e}r-Rao (BHCR) bound.
We propose a technologically feasible experiment with cold atoms to demonstrate
the quantum advantage in the simple case of two qubits.
\end{abstract}


%\tableofcontents  % optional

\markboth{Kumar Shivam, Anirudh Reddy, Joseph Samuel, Supurna Sinha}
{Entropy and Geometry of Quantum States}

\section{Introduction}
Given two quantum states, how easily can we tell them apart?
Consider for instance, gravitational wave detection which is of considerable interest in 
recent times \cite{gravwave,cramer}. Typically,
we expect a weak signal which 
produces a small change in the quantum state of the 
detector. 
The sensitivity of our instrument is determined by our 
ability to detect small changes in a quantum state. This leads to the issue of distinguishability measures
on the space of quantum states \cite{shunichi,statest, anthony, osaki, barnett}. In general, quantum states are represented by density matrices. 
In this paper, we clarify the operational meaning of two Riemannian metrics on the space of density matrices: the 
BH metric and the BKM metric.  

In fact, even in the classical domain, one encounters similar questions while considering 
drug trials, electoral predictions 
or when we compare a biased coin to a fair one. 
As the number $N$ of trials (or equivalently, the size of the sample) increases, our ability to distinguish 
between candidate probability distributions improves. Such considerations give rise 
in a natural and operational manner, to a metric on the space of probability distributions \cite{thomas}. This metric is known as
the Fisher-Rao metric and plays an important part in the theory of parameter estimation. This metric leads to the Cram\'{e}r-Rao bound
which limits the variance of any unbiased estimator.

Another example of the use of a Riemannian metric to measure distinguishability 
occurs in the theory of colours \cite{geometry,weinberg}. 
The space of colours is two dimensional (assuming normal vision) 
and one can see this on a computer screen in several graphics softwares.
The sensation of colour is determined by the relative proportion of the RGB values, which 
gives us two parameters.
The extent to which one can distinguish neighbouring colours 
is usually represented by MacAdam ellipses \cite{geometry,macadam,weinberg}, which 
are contours on the chromaticity diagram which are just 
barely distinguishable from the centre. These ellipses give us a graphical representation of an operationally defined Riemannian
metric on the space of colours. The flat metric on the Euclidean plane would be represented by circles, whose radii are everywhere the same. 
As it turns out, the metric on the space of colours is not flat and the MacAdam ellipses vary in size, orientation and eccentricity 
over the space of colours. This analogy is good to bear in mind, for we provide a similar visualization of the geometry of state 
space based on entropic considerations.

There is a subtlety here in that we started out with two distinct states (or colours) 
represented say, by points $p_1$ and $p_2$. As the second point approaches the first, we may regard
them as represented by the first point along with a tangent vector. This involves replacing a 
$\it{difference}$ by a $\it{derivative}$. One is no longer working on the space of states but on
the tangent space at a point $p_{1}$. We will refer to this as the infinitesimal limit. 
There is another limiting process involved in state discrimination: the limit of 
$N \rightarrow \infty$, where $N$ is the number of trials. We refer to this as the 
``thermodynamic" limit. Our main point in this paper is that these two limits do not
commute. If we take the infinitesimal limit first, we are led to the BH metric and the corresponding CR bound. If we choose two distinct states, no matter how small their 
separation, we find in the ``thermodynamic" limit there are quantum effects that 
give us the BKM metric as the relevant one. 
The noncommutativity of limits is the main point of this paper.

The paper is organized as follows. 
In Sec. II we review the connection between the Kullback-Leibler (KL) 
divergence and the theory of statistical inference \cite{9780511813559,thomas}. 
In Sec III we take the infinitesimal limit first and show
that this leads to the BH metric. Then we show that in the thermodynamic limit, 
the gain in discriminating power is no better than in the classical case. In Sec IV we reverse the order of limits by taking the thermodynamic limit first.
In this case, we find that as 
$N \rightarrow \infty$ our discriminating power is determined by Umegaki's quantum relative entropy between the distinct states $p_1$ and $p_2$. Taking the infinitesimal limit  leads us to the BKM metric. 
We illustrate these theoretical considerations by giving examples of the quantum advantage in the case of two qubits.
In Sec. V, we compare the quantum Cram\'er Rao bounds arising from the BH and BKM metrics. 
In Sec. VI we translate this theoretical work 
into a technologically feasible
experiment with trapped cold atoms. 
Finally, we
end the paper with some concluding remarks in Sec VII. Some calculational details are relegated to appendices A, B, and C. \ref{sec:BHmetric} computes the Bures metric as the
basis optimized Fisher-Rao metric. \ref{sec:BKMmetric} gives a simple matrix derivation of the BKM metric as the Hessian of the quantum relative entropy and \ref{sec:Geodesic} describes the 
geometry of the BKM metric and plots its geodesics. 

\section{KL Divergence as maximum likelihood}

Let us consider a biased coin for which the probability of getting a head is 
$p_H= 1/3$ and that of getting a tail is $p_T= 2/3$. Suppose we incorrectly assume 
that the coin is fair and assign probabilities $q_H= 1/2$ 
and $q_T= 1/2$ for getting a head and a tail respectively. 
The question of interest is the number of trials needed to be able to distinguish
(at a given confidence level)
between our assumed probability distribution and the measured probability distribution. 
A popular measure for distinguishing between the expected distribution and the measured distribution is 
given by the relative entropy or the KL divergence (KLD) 
which is widely used in the context of distinguishing
classical probability distributions \cite{jon}.
Let us consider $N$ independent tosses of a coin leading to a string $S=\{HTHHTHTHHTTTTT......\}$. 
What is the probability that the string is generated by the model distribution $Q=\{q,1-q\}$?
The observed frequency distribution is $P=\{p,1-p\}$. 
If there are $N_H$ heads and $N_T$ tails in a string 
then the probability of getting such a string is  $\frac{N!}{N_H!N_T!}q^{N_H}(1-q)^{N_T}$ 
which we call the likelihood function $L(N|Q)$. 
If we take the average of the logarithm of this likelihood function and use Stirling's 
approximation for large $N$ we get the following expression:
\begin{equation}
\frac{1}{N}\log{L(N|{Q)}}=-D_{KL}(P\|Q)+\frac{1}{N}\log{\frac{1}{\sqrt{2\pi Np(1-p)}}},
\label{likelihood}
\end{equation}
where $p=\frac{N_H}{N}$ and $D_{KL}(P\|Q)=p\log{\frac{p}{q}}+(1-p)\log{\frac{1-p}{1-q}}$.
The second term in (\ref{likelihood}) is due to the sub-leading term $\frac{1}{2}\log{2\pi N}$ of Stirling's approximation. 
If $D_{KL}(P\|Q)\neq 0$ then the likelihood of the string $S$ being produced by the $Q$ distribution decreases exponentially 
with $N$.
$$L(N|Q)=\frac{1}{\sqrt{2\pi Np(1-p)}} \exp{-\{ND_{KL}(P\|Q)\}}.$$ 
Thus $D_{KL}(P\|Q)$ gives us the divergence of the measured distribution from the model distribution.
The KL divergence is positive and vanishes if and only if the two distributions $P$ and $Q$ are equal. In this limit,
we find that  
the exponential divergence gives way to a power law divergence, due to the subleading term in (\ref{likelihood}).
The arguments above generalize appropriately to an arbitrary number of outcomes (instead of two) and also 
to continuous random variables.

The relative entropy (or KLD) gives an operational measure of how distinguishable two distributions are, quantified by the number of trials needed to 
distinguish two distributions at a given confidence level. However, the KLD is not a distance function on the space of probability distributions: it is not symmetric between
the distributions $P$ and $Q$. One may try to symmetrize this function, but then, the result does not satisfy the triangle inequality. However, in the infinitesimal limit, 
when $Q$ approaches $P$,
the relative entropy can be Taylor expanded to second order about $P$. The Hessian matrix does define a positive definite quadratic form at $P$ and thus  a Riemannian metric
on the space of probability distributions. For a classical probability distribution $P=\{ p_i, i=1, 2, \ldots, d \}$, the Fisher-Rao metric \cite{thomas,facchi} is given by 
\begin{equation}\label{FR}
ds^2=\sum_{i}\frac{{dp_i}^2}{p_i}
\end{equation}
and this forms the basis of classical statistical
inference and the famous $\chi$-squared test. 
The Riemannian metric then defines a distance function, based on the lengths of the shortest curves connecting any two states $P$ and $Q$.

Similar considerations also apply to the quantum case, where probability distributions are replaced by density matrices.
Consider the density matrix $\rho$ of a $d$ state system, satisfying
$\rho^\dagger=\rho$, $\text{Tr}(\rho)=1$ and $\rho>0$, where we assume $\rho$ to be {\it strictly} positive, so that we are not at 
the boundary of state space. Let 
$\boldsymbol{\lambda}=\{ \lambda^i,i=1\dotso, d^2-1\}$ be local coordinates on the space of density matrices.
Let
${\cal S}(\rho_1(\boldsymbol{\lambda_1})\|\rho_2(\boldsymbol{\lambda}))$ be a function on the space of density matrices which is positive and vanishes if and only if $\rho_2=\rho_1$ \cite{Nielsen}. 
Let us consider ${\cal S}(\rho_1(\boldsymbol{\lambda_1})\|\rho_2(\boldsymbol{\lambda}))$ as a function of its second argument.
If the states $\rho_1$ and $\rho_2$ are infinitesimally close to each other,
we can Taylor expand the relative entropy function.
\begin{equation} 
\label{exp}
{\cal S}(\rho_1\|\rho_2)={\cal S}(\rho_1\|\rho_1) + \frac{\partial {\cal S}}{\partial \lambda^i}\Delta\lambda^i +\frac{1}{2} \frac{\partial^2{\cal S}}{\partial\lambda^j\partial\lambda^i}\Delta\lambda^i\Delta\lambda^j+..
\end{equation}
Notice that ${\cal S}(\rho_1\|\rho_1)$ is zero and the second term is zero because we are doing a Taylor expansion about the minimum of the relative entropy function. 
The third term, which is second order in $\Delta \lambda$,  gives us the metric and is positive definite 
for $\Delta\lambda \neq 0$.
\begin{equation}
g_{ij}=\frac{\partial^2{\cal S}}{\partial\lambda^j\partial\lambda^i}.
\label{metric}
\end{equation}
The Hessian defines a metric {\it tensor}.
Positivity of the Hessian is guaranteed as the stationary point is {\it the absolute} minimum. The fact that density matrices do not in general commute is no obstacle to this definition.

\section{Measurements on single qubits: emergence of the Bures metric}

Let us  now consider the quantum problem of distinguishing between two states $\rho_1$ and $\rho_2$ of a qubit. Here $\rho_1=\frac{\boldsymbol{\mathds{1}+X.\sigma}}{2}$ 
plays the role of $P$ above and $\rho_2=\frac{\boldsymbol{\mathds{1}+Y.\sigma}}{2}$ that of Q. A new ingredient in the quantum problem 
is that we can choose our measurement basis. Suppose that we are given
a string of $N$ qubits all in the same state, which may be either $\rho_1$ or $\rho_2$. A possible strategy is to make projective measurements
on individual qubits, measuring the spin component in the direction $\boldsymbol{\hat m}$.  
For each choice of $\boldsymbol{{\hat m}}$ we find $p_{\pm}=\frac{\boldsymbol{1\pm X.{\hat m}}}{2}$ and $q_{\pm}=\frac{\boldsymbol{1\pm Y.{\hat m}}}{2}$ and we can compute the KL-Divergence 
or the classical relative entropy of the two distributions as :

\begin{equation}\label{kldm}
S_{\boldsymbol{m}}(\rho_{1}\|\rho_{2})=p_{+}\log{\frac{p_{+}}{q_{+}}}+p_{-}\log{\frac{p_{-}}{q_{-}}}.
\end{equation}
We will now choose  $\boldsymbol{{\hat m}}$ in such a way as to maximize our discriminating power {\it i.e} $S_{\boldsymbol{m}}(\rho_{1}\|\rho_{2})$. This gives us, 

\begin{equation}\label{opt1}
\delta S_{\boldsymbol{m}}=\frac{\partial S}{\partial \boldsymbol{{\hat m}}}\delta \boldsymbol{{\hat m}}=\lambda \delta \boldsymbol{{\hat m}},
\end{equation}
which can be rewritten as 

\begin{equation}\label{opt2}
\frac{\partial S_{\boldsymbol{m}}}{\partial a_1}\boldsymbol{X}+\frac{\partial S_{\boldsymbol{m}}}{\partial a_2}\boldsymbol{Y}=\lambda \delta \boldsymbol{{\hat m}},
\end{equation}
where $a_1=\boldsymbol{{\hat m}.X}$ and $a_2=\boldsymbol{{\hat m}.Y}$. Since $\delta S_{\boldsymbol{m}}$ is a linear combination of $\boldsymbol{X}$ and $\boldsymbol{Y}$ we find 
that $\boldsymbol{{\hat m}}$ must lie in the plane containing $\boldsymbol{X}$ and $\boldsymbol{Y}$, as shown in Fig. 1. Without loss of generality, we can suppose this to be the $x-z$ plane,
so that $X_2=Y_2=\hat{m}_2=0$. 
We can replace $\boldsymbol{{\hat m}}=(\cos{\beta},0,\sin{\beta})$ by the angle $\beta$, which gives us $p_{\pm}=\frac{1}{2} (1\pm r_1\cos{\beta}) $ and $q_{\pm}=\frac{1}{2}(1\pm r_2\cos{(\theta+\beta)})$, where $r_{1}=|\boldsymbol{X}|$ and $r_{2}=|\boldsymbol{Y}|$. Plotting $S(\beta)$ (Fig. 2), we find that the maximum distinguishability is 
attained at $\beta=\beta^*$ . This is clearly the most advantageous choice of 
$\beta$. The value of $S_{\boldsymbol{m}}$ at the maximum is denoted by $S^*(r_1,r_2,\theta)=S_{\boldsymbol{m}}(r_1,r_2,\theta,\beta^*(r_1,r_2,\theta))$. $S^*(r_1,r_2,\theta)$ 
gives us the optimal choice for state discrimination when we measure qubits, one at a time. As we can see in Fig. 2, $S^*(r_1,r_2,\theta)$ is 
never more than 
Umegaki's quantum relative entropy\cite{umegaki}.
\begin{equation}
S(\rho_1(\boldsymbol{\lambda_1})\|\rho_2(\boldsymbol{\lambda}))= \text{Tr}[\rho_1 \log\rho_1-\rho_1\log\rho_2].
\label{umegaki}
\end{equation}

Equality between $S^*(\rho_1\|\rho_2)$ and $S(\rho_1\|\rho_2)$ happens  
if and only if $[\rho_1,\rho_2]=0\ (\theta=0,\pi, 2\pi\approx 0)$ [See Fig. 3] {\it i.e} when the two density matrices commute with each other. 

\begin{figure}[h!]
	\begin{center}
		\includegraphics[width=0.6\textwidth]{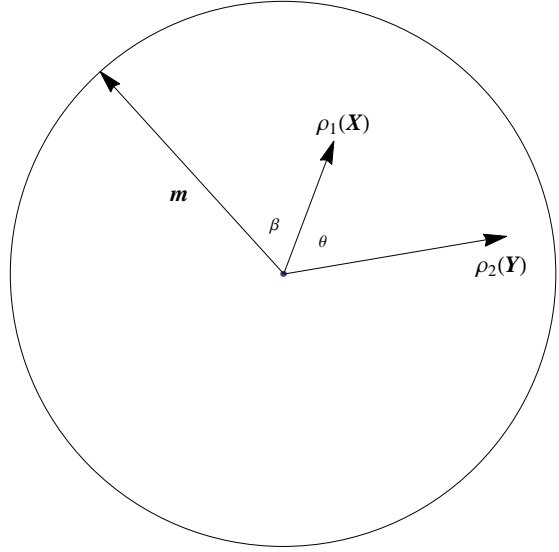}
		\caption{The figure shows the measurement direction $\boldsymbol{{\hat m}}$ and directions 
			$\boldsymbol{X}$ and $\boldsymbol{Y}$ corresponding to the density matrices $\rho_1(\boldsymbol{X})$ and 
			$\rho_2(\boldsymbol{Y})$ respectively.}
	\end{center}
\end{figure}

\begin{figure}[h!]
	\begin{center}
		\includegraphics[width=0.8\textwidth]{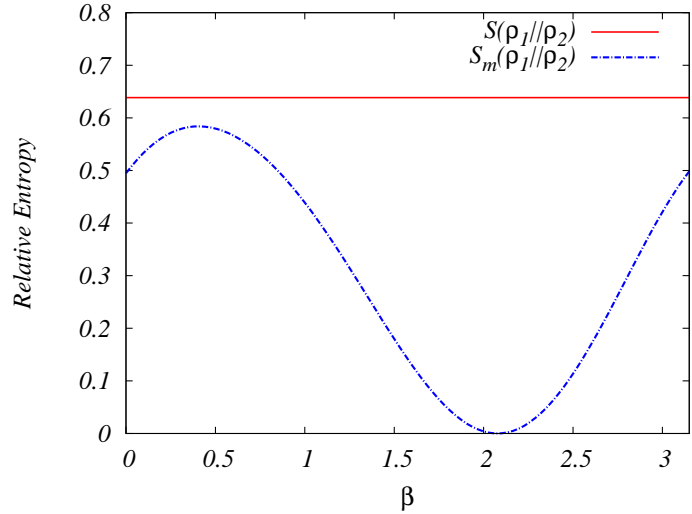}
		\caption{The relative entropy between two density matrices 
			as a function of the measurement basis parametrized by $\beta$. The maximum occurs at $\beta^*=0.41$. 
			For comparison we also show in red the horizontal line
			representing Umgegaki's relative entropy (\ref{umegaki}).
		}\end{center}
	\end{figure}
	
	We now take the infinitesimal limit and replace $(\rho_{1},\rho_{2})$ by $(\rho, d\rho)$ and represent $\rho$ by $(r,\theta)$ and $d\rho$ by $(dr,d\theta)$.
	$dp_+$ and $dp_-$ are:
	\begin{equation} \label{dp}
	\left.\begin{aligned}
	dp_+=\frac{\cos\beta dr-r\sin\beta d\theta}{2}, \\
	dp_-=\frac{r\sin\beta d\theta-\cos\beta dr}{2}.
	\end{aligned}
	\right\}
	\qquad
	\end{equation}
	
	\begin{figure}[h!]
		\begin{center}
			\includegraphics[width=0.8\textwidth]{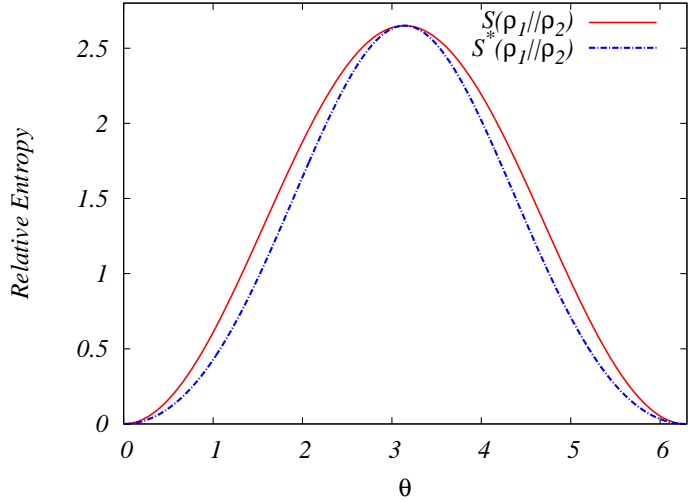}
			\caption{The figure shows the quantum relative entropy and the classical
				relative entropy as a function of $\theta$, for $r_1=0.9$ and $r_2=0.9$. Note that the quantum relative entropy
				in general exceeds the classical one. This difference is what we call the quantum advantage, which can be exploited to beat the BHCR bound. The quantum relative entropy equals the classical one only for $\theta =0,\pi,2\pi\approx 0$.}
		\end{center}
	\end{figure}
	
	Considering the classical relative entropy (\ref{kldm}) 
	between infinitesimally separated states and doing a Taylor expansion, 
	gives us the Fisher-Rao metric \eqref{FR}, which is given by
	\begin{equation}
	\label{frmetric}
	ds^2=\frac{dp_+^2}{p_+}+\frac{dp_-^2}{p_-}.
	\end{equation}
	Maximising \eqref{frmetric} with respect to the measurement basis for fixed $\rho_{1}$ and $d\rho$ gives us an expression for the metric (see \ref{sec:BHmetric} for a detailed calculation)
	\begin{equation} \label{bs}
	ds^2=\frac{dr^2}{1-r^2}+r^2d\theta^2.
	\end{equation}
	Returning to three dimensions using spherical symmetry we 
	get an expression for the metric
	\begin{equation} \label{bs1}
	ds^2=\frac{dr^2}{1-r^2}+r^2(d\theta^2+\sin^2{\theta}d\phi^2).
	\end{equation}
	In the above derivation, we have defined the 
	distinguishability metric in the tangent space by optimising over all
	measurement bases. The metric we arrive at is 
	the BH metric (Ref. \cite{bures,helstrom,caves,dittmann,Uhlmann1992,ericsson} and references therein), which was introduced by Bures \cite{bures} 
	from a purely mathematical point of view. Its relevance to quantum state discrimination was elucidated by Helstrom\cite{helstrom}.
	It plays the role of the Fisher-Rao metric in quantum physics, {\it if one restricts oneself to measuring one qubit at a time}.

	More generally, the BH metric is defined as follows\cite{helstrom, erc1, erc2}. 
	Let $d\rho$ be a tangent vector at $\rho$. 
	Consider the equation for the unknown $L$:
	\begin{equation}
	d\rho=\frac{1}{2}\{\rho,L\}
	\label{ldef}
	\end{equation}
	This linear equation defines the symmetric logarithmic derivative $L$ uniquely.
	Optimising the Fisher-Rao metric \eqref{FR},
	$g_{FR}(\rho,d\rho)=\sum_i{{\bra{i}\rho\ket{i}^{-1}}{\bra{i}d\rho\ket{i}^{2}}}$
	over all choices of orthonormal bases $b=\{\ket{i}, i=1,2,3 ....d \}$ we find that \cite{helstrom} 
	(A) the optimal choice is given by the basis $b^*$ which diagonalizes $L$ and
	(B) that the optimal value is given by 
	$$ g_{BH}(\rho,d\rho) = Tr[\rho L L]$$ which is defined as the Bures metric.
	The discussion above is general and applicable to a $d$ state system. 
	
	We now take the thermodynamic limit.
	Consider $N$ qubits with the state $\rho^{\otimes N}$. We will show that
	\begin{equation}
	\frac{1}{N}g_{BH}(\rho^{\otimes N},d \rho^{\otimes N})=g_{BH}(\rho,d\rho)
	\label{additive}
	\end{equation}
	The proof is by induction. For $N=1$ (\ref{additive}) is an identity. Assuming (\ref{additive}) for $N-1$, we note that 
	$$d\rho^{\otimes N}=d(\rho^{\otimes N-1}{\otimes}\rho)=d\rho^{\otimes N-1}\otimes \rho+\rho^{\otimes N-1}\otimes d\rho,$$
	and that 
	$$L_N=L_{N-1}\otimes \unit+\unit\otimes L$$
	uniquely solves (\ref{ldef}).
	
	Computing $g_{BH}(\rho^{\otimes N},d \rho^{\otimes N})=Tr[\rho^{\otimes N} L_N L_N]$ and using the fact that (\ref{ldef}) implies $Tr[\rho L]=0$ we arrive at (\ref{additive}).
	The optimized Fisher-Rao metric has the same discriminating power (per qubit) for $N$ qubits as for a 
	single qubit. This is exactly as in the classical case. This holds true in the limit $N\rightarrow \infty$. There is no quantum advantage. Note that in the above, we have taken the infinitesimal limit
	first. We will see that taking the thermodynamic limit first leads to an entirely different picture. 
	
	\section{Quantum advantage : Measurements on Multiple Qubits}
	Let us now take the ``thermodynamic'' limit of large $N$ first. Given
	$N$  qubits, which may be a state $\rho_1^{\otimes N}$ or 
	$\rho_2^{\otimes N}$ we can choose a measurement basis in the Hilbert space
	${\cal H}^{\otimes N}$. The optimization over measurement bases is now over an enlarged set.
	Earlier we were restricted to bases of the form $b^{\otimes N}$ which are separable in the Hilbert space ${\cal H}^{\otimes N}$. 
	We now have the freedom to include entangled bases and this implies
	
	\begin{equation}
	\frac{S^*(\rho_1^{\otimes N}\|\rho_2^{\otimes N})}{N}\ge S^*(\rho_1\|\rho_2).
	\label{quadvantage}
	\end{equation}
	In fact\cite{vedralrmp}, no matter how small the separation between the distinct states 
	$\rho_{1}$ and $\rho_{2}$, as $N\rightarrow\infty$, $\frac{1}{N} S^*(\rho_1^{\otimes N}\|\rho_2^{\otimes N}) \rightarrow S(\rho_1\|\rho_2)$,
	where $S(\rho_1\|\rho_2)$ is Umegaki's quantum relative entropy.  As we see in Fig. 3, 
	this is greater than or equal to the classical relative entropy, so the appropriate 
	relative entropy to use in the thermodynamic limit is Umegaki's relative entropy. 
	
	If we now take the infinitesimal limit as $\rho_2\rightarrow\rho_1$,
	we effectively pass from the quantum relative entropy to a Riemannian metric defined as the Hessian of the quantum relative entropy. The form of this metric
	in the case of a qubit is (see \ref{sec:BKMmetric})
	
	\begin{equation} \label{me}
	\boxed{g_{ij}=\frac{\partial^2S}{\partial x^i\partial x^j}=C(r)\frac{x^ix^j}{r^2} + D(r)\{\delta_{ij}-\frac{x^ix^j}{r^2}\}},
	\end{equation}
	where $C(r)=\frac{1}{1-r^2}$, $D(r)=\frac{1}{2r}\log\left(\frac{1+r}{1-r}\right)$ and $r=|\boldsymbol{Y}|$.
	
	%--------------------------------------------------------------------------------------------------------------------------------
	The corresponding line element is given in polar coordinates by: 
	\begin{equation} \label{linel}
	ds^2=\frac{dr^2}{1-r^2}+\left[\frac{r}{2}\log\left(\frac{1+r}{1-r}\right)\right]{(d\theta^2+\sin^2\theta d\phi^2)}.
	\end{equation}
	This metric has been discussed earlier by Bogoliubov, Kubo and Mori (BKM) in the context of statistical mechanical fluctuations \cite{Petz,km,BALIAN}. 
	We refer to it as the BKM metric. For a  discussion on the geometry of the BKM metric see \ref{sec:Geodesic}.
	
	To illustrate the quantum advantage that comes from grouping qubits before measuring them, 
	we numerically study an example for $N=2$ and $\rho_1,\ \rho_2$ distinct and well separated. 
	The quantum state of the combined system is now given by $\tilde{\rho}=\rho\otimes \rho$,
	where $\rho$ can refer to either $\rho_1$ or $\rho_2$. In choosing a measurement basis to distinguish $\tilde{\rho_1}$ from $\tilde{\rho_2}$,
	we now have the additional advantage that we can choose bases which are not separable. This extra freedom gives us the quantum advantage
	which comes from entanglement.
	For example, let us choose $(r_1,r_2,\theta)=(0.9,0.5,\pi/2)$ so that 
	$\boldsymbol{X}=\{r_1,0,0\},\ \boldsymbol{Y}=\{r_2/\sqrt{2},0,r_2/\sqrt{2}\}$ and the direction $\boldsymbol{\hat{m}}$ in the $x$-$z$ plane
	$\boldsymbol{\hat{m}}=\{\cos{\beta},0,\sin{\beta}\}$. Let the corresponding 1-qubit basis
	which diagonalizes $\hat{m}.\boldsymbol{\sigma}$ be $\ket{+},\ket{-}$. We now construct the non separable basis
	$|b_1\rangle=\frac{|{+-}\rangle+|{-+}\rangle}{\sqrt{2}}$,
	$|b_2\rangle=\frac{|{+-}\rangle-|{-+}\rangle}{\sqrt{2}}$,
	$|b_3\rangle=|{++}\rangle$
	and $|b_4\rangle=|{--}\rangle$.
	Note that two of these basis states are maximally entangled Bell states and two are completely separable
	(Curiously, using all basis states as Bell states leads to no improvement over the separable states). 
	We numerically compute the relative entropy and optimize over $\beta$. This leads to an improvement over
	measurements conducted on one qubit at a time. The improvement is seen in the value of the relative entropy per qubit,
	which increases from $0.5839$ in the one qubit strategy to $0.5856$ in the two qubit strategy.
	
	In fact, this number can be further
	improved. By numerical Monte-Carlo searching, we have found bases (which don't have the clean form above) which yield
	a relative entropy of $0.5863$ per qubit. Our Monte-Carlo search is simplified by the observation that one can by a unitary transformation 
	bring any two states described by $\boldsymbol{X}$ and $\boldsymbol{Y}$ to the $x$-$z$ plane of the Bloch ball, so that we are working over the real numbers
	rather than complex numbers. Over the reals, unitary matrices are orthogonal matrices. We start with an initial basis
	in the four dimensional real Hilbert space of the composite system and then rotate the basis by a random orthogonal matrix
	close to the identity. We then compute the relative entropy using the new basis and accept the move if the new basis
	has a larger relative entropy and reject it otherwise. This gives us a monotonic rise in the relative entropy and 
	drives us towards the optimal basis in the two qubit Hilbert space. 
	
	The method extends easily to three qubits and 
	more although the searches are more time consuming.
	We have numerically observed that measuring three qubits at a time results in a further improvement over the two qubit
	measurement strategy. However, this number (0.5880) still falls short of the quantum relative entropy which is $0.6385$.
	The classically optimized relative entropy $S^{*}_N$ for $N$ qubits considered as a single system satisfies the inequality
	$\frac{1}{N}S^{*}_N \leq S_Q$ \cite{vedralrmp} where $S_Q$ is the quantum relative entropy.
	As $N\rightarrow\infty$ the inequality is saturated.   
	Thus the gap between the classically optimized relative entropy and
	the quantum relative entropy (Fig. 2 and Fig. 3) 
	progressively reduces as one increases the number of qubits measured at a time.
	
	\section{QUANTUM CRAM\'ER RAO BOUNDS}
	
	As we have seen, 
	the quantum relative entropy 
	leads us to a metric (the BKM metric) on the tangent space.
	We notice (see Fig. 2 and Fig. 3) that the quantum relative entropy dominates over the classically optimized relative
	entropy computed in Sec III : $S(\rho_1\|\rho_2)\ge S^*(\rho_1\|\rho_2)$\cite{vedralrmp}. This implies that $g_{BKM}(v,v)\ge g_{BH}(v,v)$ for all tangent vectors $v$.
	This can be explicitly seen by comparing Eq. (\ref{linel})  
	with Eq. (\ref{bs1}) and noting that $r/2\log{[(1+r)/(1-r)]}\ge r^2$.
	This means that the BKM metric is more {\it discriminating} than the BH metric in the sense that distances are larger.
	Figure 4 shows a graphical representation of the geometry of state space as given by the BH metric (white ellipses) and
	the BKM metric \eqref{linel} (in black). Geometrically the unit sphere of the BKM metric is contained within 
	the unit sphere of the BH metric (Fig. 4). 
	
	The higher discrimination of the BKM metric over the BH metric translates
	into a {\it less stringent} Cram\'{e}r-Rao bound, since the bound is based on the inverse of the metric. 
	Let $X$ be an unbiased estimator for a parameter $\theta$.
	Then the variance $V=\text{Tr}[\rho X X]-(\text{Tr}[\rho X ])^2$  has to satisfy $V\ge \frac{1}{g(v,v)}$.
	This is the well known Cram\'er-Rao bound.

	\begin{figure}\label{macadam}
		\begin{center}
			\includegraphics[width=1.0\textwidth]{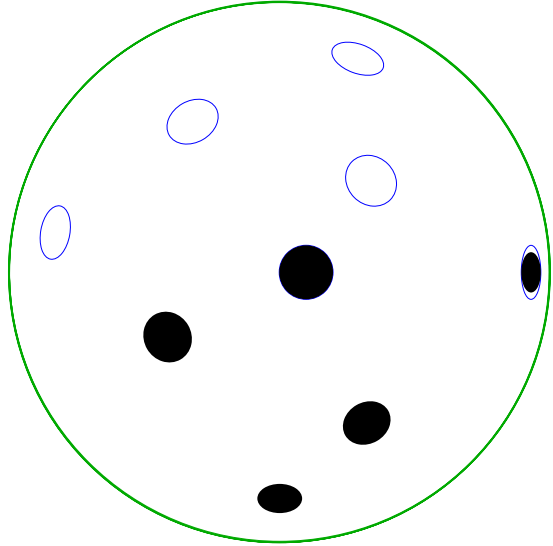}
			\caption{The figure represents the geometry of the qubit state space as given by 
				the BKM metric (black ellipses in the lower half) and the BH metric (white ellipses with blue (online) boundaries  in the upper half). The figure shows a two dimensional slice of the 
				three dimensional qubit state space. The geometry is invariant under rotations due to the unitary symmetry of the state space. Note
				that the ellipticity increases near the boundary of the state space. The ellipse on the right shows both BH and BKM metrics superposed. 
				Note that the black BKM ellipse is {\it inside} the white BH ellipse and the white region represents the quantum advantage.}
		\end{center}
	\end{figure}
	
	In order to bring out this point, we propose an experimentally realizable strategy that 
	exploits the dominance of the BKM metric over the BH metric. 
	As we have seen in Sec III, restricting to measurements of one qubit at a time we find that 
	the BH metric sets the limit on state discrimination \cite{erc1, erc2}. However, as we 
	have seen in Sec IV, that we can beat this limit by measuring multiple qubits at a time.

	\section{Proposed Experimental Realization}
	The strategy described above can be experimentally realized with current technology using cold atoms in traps.
	Experimental realizations of the quantum advantage are within reach. 
	There have been studies involving measurements for quantum state discrimination \cite{anthony, osaki, barnett}, 
	where the upper limit of the state distinguishability is set by the BHCR bound. In order to exploit the quantum advantage discussed 
	here and go beyond the BHCR bound, we need to measure in an entangled basis of the two qubit system. 
	The entangled basis $\ket{b_i}$ mentioned here, is related to the separable 
	basis $|++\rangle,\ |+-\rangle,\ |-+\rangle,\ |--\rangle$ by a unitary transformation 
	$U$ in the four dimensional Hilbert space. 
	One can equivalently apply $U$ to the separable state $\tilde{\rho}=\rho\otimes\rho$. This creates an entangled 
	state $U^\dagger \tilde{\rho} U$, which can then be measured in the separable basis using a 
	projective measurement. Consider a pair of qubits subject to the Hamiltonian 
	
	\begin{equation}
	H=\vec{\sigma_1}.\vec{B_1}+\vec{\sigma_2}.\vec{B_2}+ J(t) \vec{\sigma_1}.\vec{\sigma_2},
	\label{ham2}
	\end{equation}
	which is a standard Heisenberg Hamiltonian for spins. 
	This Hamiltonian evolution produces the unitary transformation $U$ for a suitable choice of 
	$J(t)$. 
	
	This entangling unitary transformation $U$
	is the square root of the SWAP operation $U=\sqrt{SWAP}$. $U$ has already been experimentally
	realized in \cite{phillips} by creating a system in the laboratory subject to the Hamiltonian (\ref{ham2}).  The method used in \cite{phillips} is to load ${}^{87}Rb$ atoms in pairs
	into an array of double well potentials. The experimenters have control over all the parameters in the Hamiltonian. 
	They can generate the transformation $U$ at will by using a $\pi/4$ pulse for $J(t)$ by using 
	radio frequency, site selective pulses to address the qubits in pairs (See Table 1 of \cite{phillips}),
	thus effecting  the entangling unitary transformation $U$. What remains to be done to implement
	our proposal is to projectively measure each of the qubits separately and thus achieve 
	a violation of the BHCR bound in distinguishing states.

	\section{Conclusion}
	
	The main goal of this paper is to draw attention to a noncommutativity of 
	limits in the context of quantum state discrimination. In particular, there are two limits --- one which we call the 
	``thermodynamic" limit (of N, the number of realizations going to infinity)
	and the infinitesimal limit (of the separation of states tending to zero) --- which do not commute in the quantum case.
	We show that taking the infinitesimal limit first leads to the BH metric. In contrast, taking the ``thermodynamic' limit first 
	leads to the BKM metric. The lack of commutation of limits is a purely quantum phenomenon with no classical counterpart. 
	We have explicitly shown by numerical methods, that one can make use of this lack of commutation of limits 
	to make use of quantum entanglement to get an advantage in state 
	discrimination.
	
	Questions addressed here were raised but not fully answered in an early paper of Peres and Wootters \cite{peres}. At that time it was not
	fully clear whether there was  a one qubit strategy which could compete with the multiqubit strategy. Subsequent work using the machinery 
	of $C^*$ algebras has made it clear \cite{hiai,vedralrmp} that the best one qubit strategy is inferior to the multiqubit strategy. As $N$ increases
	we approach the bound set by the BKM metric. Thus the quantum Cram\'{e}r-Rao bound set by the BKM metric can be approached but not surpassed. In contrast, the 
	BHCR bound can be surpassed, as we have seen in Sec VI. 
	
	We have worked out the geodesics of the BKM metric and plotted them numerically. We have noticed that any two points are connected
	by a unique geodesic. The BKM metric leads to a distance function on the state space that emerges naturally from entropic and geometric
	considerations. In working out the geodesics, it is easily seen analytically 
	that the geodesics approach the boundary of the state space at right angles. However, this approach is logarithmically slow and is not
	apparent in Fig. 6. The form of the geodesics on state space is reminiscent of the geodesics of the Poincar\'{e} metric which also meet
	the boundary at right angles. However, there are serious differences. While both metrics have negative curvature, the Poincar\'{e} metric has a {\it constant} negative curvature, unlike the BKM metric that has a varying curvature, which diverges logarithmically at the boundary.
	It is natural to ask if this is a genuine singularity or one caused by our choice of coordinates. It is easily seen that the singularity is genuine. Consider a radial geodesic starting from $r=r_0$ and reaching the boundary at $r=1$. Its length is given by $\int_{r_0}^1 dr/\sqrt{1-r^2}=\pi/2-\arcsin{r_0}$, 
	which is finite. So the geodesic reaches the singularity of $R$ in a finite distance. Since the length of the geodesic and the scalar curvature are 
	independent of coordinates, it follows that the singularity is genuine and not an artifact of the 
	coordinate system. The divergence of the metric as one approaches $r=1$ has a physical interpretation.  
	It means that pure states offer a much larger quantum advantage than mixed states. In fact, quantum advantage diverges logarithmically as we approach the pure state limit. 
	Conversely, even a small corruption of the purity of quantum states 
	will seriously undermine our ability to distinguish between them.
	
	From the statistical physics perspective, 
	the BKM metric can be interpreted as a thermodynamic susceptibility of a quantum state $\rho$ (viewed as a Gibbs state for the Hamiltonian $H=-(1/\beta)\log{\rho}$), to perturbations. The Gibbs state is the state that maximizes its entropy subject to an energy constraint. However, in statistical physics, a system makes spontaneous excursions to neighbouring lower entropy states. The size of these fluctuations is determined by the Hessian of the entropy function and thus related to the susceptibility.
	
	In the existing literature\cite{petznew,petznew2,hasenew,jencova,grasselli} researchers have discussed the BKM and other Riemannian metrics on the quantum state space but have mainly focussed on the geometrical and mathematical aspects of the metric. In the context of quantum metrology\cite{a,b} the idea that a quantum procedure leads to an improved sensitivity in parameter estimation compared to its classical counterpart has been explored.
	
	We go beyond earlier studies in suggesting physical and statistical mechanical interpretations of the geometry 
	and an experimental proposal demonstrating the use of entanglement as a resource. Such an experimental demonstration would operationally 
	bring out a subtle aspect of quantum information. We hope to interest experimental colleagues in this endeavour.

\section{Acknowledgement}
It is a pleasure to thank Nomaan Ahmed, Giuseppe Marmo, Saverio Pascazio and Rafael Sorkin for discussions. We also acknowledge Saptarishi Chaudhuri and Sanjukta Roy for discussions on possible experimental realizations.

\appendix
\label{appendix}
\section{BH METRIC FOR A QUBIT}
\label{sec:BHmetric}
The Fisher-Rao metric is given by
\begin{equation}
ds^2=\frac{dp_+^2}{p_+}+\frac{dp_-^2}{p_-} \nonumber.
\end{equation}
%\pagebreak
Substituting $dp_+$, $dp_-$, $p_+$ and $p_-$, from \eqref{dp}, we get
\begin{equation} \label{dss}
ds^2=\frac{\left(dr-r\tan\beta d\theta\right)^2}{1-r^2+\tan^2\beta}.
\end{equation}
Keeping $r$, $dr$, $d\theta$ fixed and optimising with respect to $\beta$ we find 
\begin{equation}
\tan\beta^{*}=-\frac{r\left(1-r^2\right)}{dr/d\theta}.
\end{equation}
Substituting $\tan\beta^{*}$ in \eqref{dss} we get the expression for the metric
\begin{equation} \label{bs}
ds^2=\frac{dr^2}{1-r^2}+r^2d\theta^2.
\end{equation}

\section{BKM METRIC FOR A QUBIT}
\label{sec:BKMmetric}
Consider two mixed states $\rho_1$ and $\rho_2$ of a two level quantum system commonly referred to as a qubit. These can be written as
$\rho_1=\frac{\boldsymbol{\mathds{1}+X.\sigma}}{2}$ and $\rho_2=\frac{\boldsymbol{\mathds{1}+Y.\sigma}}{2}$
where $|\boldsymbol{X}|$ and $|\boldsymbol{Y}|$ $<$ 1. $\boldsymbol{X}$ and $\boldsymbol{Y}$ are three dimensional vectors with components
$x^i$ and $y^i$. 
The relative entropy function can be written as follows:
\begin{eqnarray}
S(\rho_1\|\rho_2)&=&\text{Tr}\left[\left(\frac{\boldsymbol{\mathds{1}+X.\sigma}}{2}\right)\log\left(\frac{\boldsymbol{\mathds{1}+X.\sigma}}{2}\right)\right] \nonumber \\
&&-\text{Tr}\left[\left(\frac{\boldsymbol{\mathds{1}+X.\sigma}}{2}\right)\log\left(\frac{\boldsymbol{\mathds{1}+Y.\sigma}}{2}\right)\right].
\end{eqnarray}
We can use the power series expansion of $\log(\boldsymbol{\mathds{1}+Y.\sigma})$ to evaluate the trace of the above expression.
\begin{equation}
\log(\boldsymbol{\mathds{1}+Y.\sigma})=\underbrace{\left(\sum_{m=0}^{\infty}\frac{|\boldsymbol{Y}|^{2m+1}}{2m+1}\right)}_{f_o(|\boldsymbol{Y}|)}\frac{\boldsymbol{Y.\sigma}}{|\boldsymbol{Y}|}+\underbrace{\left(\sum_{n=0}^{\infty}\frac{|\boldsymbol{Y}|^{2n}}{2n}\right)}_{f_e(|\boldsymbol{Y}|)}\mathds{1},
\end{equation}
where $f_o(|\boldsymbol{Y}|)$ and $f_e(|\boldsymbol{Y}|)$ are respectively the odd and even parts of the function $f(r)=\log{(1+r)}$ . Notice that the odd part of the expansion is traceless. Making use of the above expansion we can express $S(\rho_1\|\rho_2)$ as follows
\begin{equation}
S(\rho_1\|\rho_2)=S(\boldsymbol{X}\|\boldsymbol{Y})=f_e(|\boldsymbol{Y}|)-\frac{f_o(|\boldsymbol{Y}|)}{|\boldsymbol{Y}|}(\boldsymbol{X}.\boldsymbol{Y}).
\end{equation}
In order to compute the Hessian of $S(\rho_1\|\rho_2)$ we 
compute the second derivative $\frac{\partial^2 S}{\partial y^i \partial y^j}$
with respect to $y^j$ and then set $y^i=x^i$ and obtain the following metric \cite{geometry}:

\begin{equation} \label{me}
g_{ij}=\frac{\partial^2S}{\partial x^i\partial x^j}=C(r)\frac{x^ix^j}{r^2} + D(r)\{\delta_{ij}-\frac{x^ix^j}{r^2}\},
\end{equation}
where $C(r)=\frac{1}{1-r^2}$, $D(r)=\frac{1}{2r}\log\left(\frac{1+r}{1-r}\right)$ and $r=|\boldsymbol{Y}|$.

\section{GEOMETRY OF THE BKM METRIC}
\label{sec:Geodesic}
The scalar curvature $R$ of the BKM metric is given by:
\begin{equation} \label{ka}
R=\frac{4r^2-4r(1+r^2)\log(\frac{1+r}{1-r})+(1+2r^2-3r^4)[\log(\frac{1+r}{1-r})]^2}{2r^2(1-r^2)[\log(\frac{1+r}{1-r})]^2}.
\end{equation}
\begin{figure}[h!]
	\begin{center}
		\includegraphics[width=0.8\textwidth]{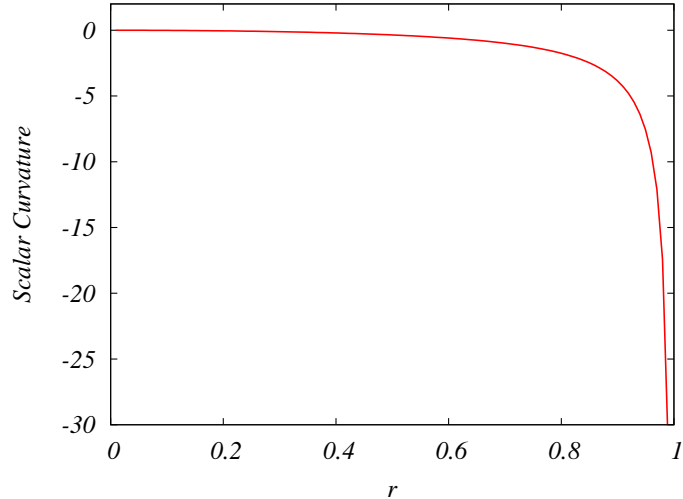}
		\caption{The scalar curvature \eqref{ka} for the 
			metric displayed in \eqref{me} as a function of $r$, distance from the centre of the Bloch sphere.}
		\label{1}
	\end{center}
\end{figure}
As we can see from Fig. 5, the metric has  negative scalar curvature and therefore the geodesics (Fig. 6) cannot cross 
more than once. It follows therefore that any two states, 
are connected by an unique geodesic. The length of this geodesic gives us a distance on the space of states. This has all the properties expected of a distance function: it is symmetric, strictly positive between distinct points and satisfies the triangle inequality.
The scalar curvature is zero near the origin
and diverges logarithmically to minus infinity as $r$ goes to unity. The geodesics of this metric are easily worked out from classical mechanics.
The metric has spherical
symmetry, because the quantum state space is invariant under unitary transformations.   

Setting $r=\sin{\alpha}$, we rewrite the metric as
\begin{equation}
ds^2=d\alpha^2
+F(\alpha) \left(d\theta^2+\sin^2(\theta)d\phi^2\right),
\end{equation}
where
$F(\alpha)=\frac{\sin\alpha}{2}\log\left[ \frac{1+\sin\alpha}{1-\sin\alpha} \right]$.
Because of the spherical symmetry,
there is a conserved angular momentum vector $\vec{J}$ and thus the geodesics lie in the plane 
perpendicular to $\vec{J}$. 
Thus we can confine our calculations to a plane, 
reducing the form of the metric to
\begin{equation}
ds^2=d\alpha^2+F(\alpha) \left(d\phi^2\right),
\end{equation}
where we have set $\theta=\frac{\pi}{2}$.
The Lagrangian of the classical mechanical system is 
\begin{equation}
L=\frac{1}{2}\left({\dot{\alpha}}^2+F(\alpha){\dot{\phi}}^2\right).
\end{equation}
The constants of motion for this problem are the energy and the angular momentum, which are given by
\begin{equation}
E=\frac{1}{2}\left({\dot{\alpha}}^2+F(\alpha){\dot{\phi}}^2\right), \
P_\phi=J=\frac{\partial L}{\partial {\dot{\phi}}}=F(\alpha)\dot{\phi}.
\end{equation}
Using the above equations we solve for $\dot{\alpha}$ and $\dot{\phi}$. Our numerical solution 
gives us the geodesics of interest. A typical geodesic is displayed in Fig. 6. Given any two points in the state space
(for example the red dots of Fig. 6), the length of the unique geodesic \cite{negcurv} connecting them gives us a distance function.
This is very similar in spirit to a construction of Wootters \cite{wootters}, who introduced a metric based on distinguishability
for {\it pure} states and used this to define a metric on pure states, which ultimately yielded the Fubini-Study metric. This work can be viewed
as an application of Wootters' idea to mixed states.

\begin{figure}[h!]
	\begin{center}
		\includegraphics[width=0.8\textwidth]{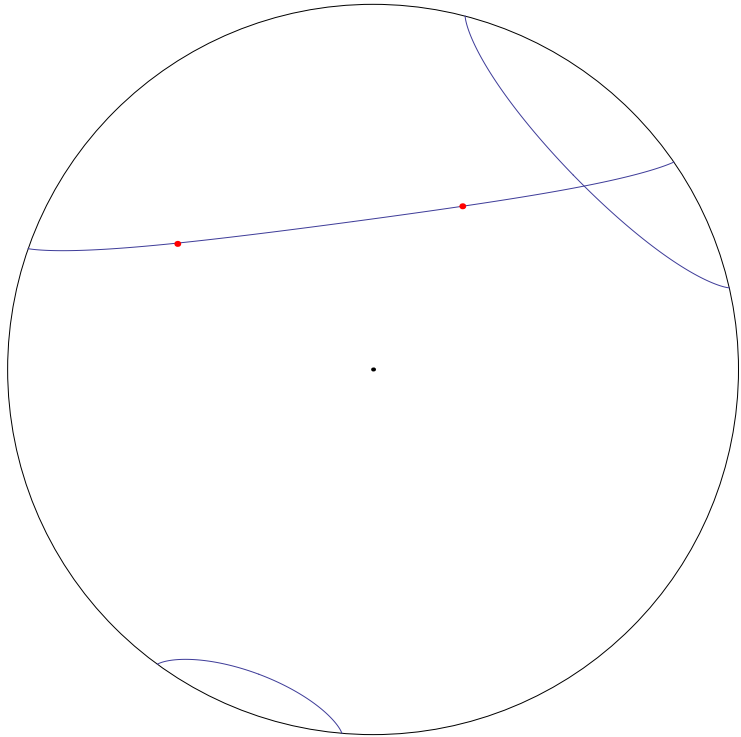}
		\caption{The figure shows a geodesic connecting two 
			typical quantum states, 
			indicated by two dots on the Bloch ball. 
			Two more geodesics are shown with different values of $J=|\vec J|$. 
			We also show geodesics crossing each other once. As explained in the text, the metric on quantum state space has negative curvature and
			so geodesics cannot cross more than once.}
		\label{2}
	\end{center}
\end{figure}
\pagebreak

\bibliographystyle{unsrt}
%\bibliography{metricreferences.bib}

\end{document}